\documentclass[final,3p,times,twocolumn]{elsarticle}

\usepackage{amssymb}
\usepackage{amsmath}
\usepackage{hyphenat}
\usepackage{color}

\journal{}

\begin{document}

\begin{frontmatter}



\title{Prospects of Passive Radio Detection of a Subsurface Ocean on Europa with a Lander}

\author[label1]{Andrew Romero-Wolf\corref{cor1}}
\ead{Andrew.Romero-Wolf@jpl.nasa.gov}
\cortext[cor1]{Corresponding Author. Tel +1 6263909060}
\author[label2]{Dustin M. Schroeder}
\author[label1]{Paul Ries}
\author[label1]{Bruce G. Bills}
\author[label1]{Charles Naudet}
\author[label1]{Bryan R. Scott}
\author[label1]{Robert Treuhaft}
\author[label1]{Steve Vance}
\address[label1]{Jet Propulsion Laboratory, California Institute of Technology, 4800 Oak Grove Drive, Pasadena, CA 91101}
\address[label2]{Department of Geophysics, Stanford University, Stanford, CA 94305}


\begin{abstract}
We estimate the sensitivity of a lander-based instrument for the passive radio detection of a subsurface ocean beneath the ice shell of Europa, expected to be between 3~km - 30~km thick, using Jupiter's decametric radiation. A passive technique was previously studied for an orbiter. Using passive detection in a lander platform provides a point measurement with significant improvements due to largely reduced losses from surface roughness effects, longer integration times, and diminished dispersion due to ionospheric effects allowing operation at lower frequencies and a wider band. A passive sounder on-board a lander provides a low resource instrument sensitive to subsurface ocean at Europa up to depths of 6.9~km for high loss ice (16 dB/km two-way attenuation rate) and 69~km for pure ice (1.6 dB/km).
\end{abstract}


\end{frontmatter}



%
%
%
%
%
%
%
%
%

In a previous study, Romero-Wolf et al. 2015 explored the sensitivity to detecting a subsurface ocean beneath the ice shell of Europa using Jupiter's decametric (DAM) radio emission with an orbiter. The technique exploits the fact that Jupiter's DAM, when active, is the brightest object in the 0.3-40~MHz band in Europa's sky. Surface and subsurface reflections of this signal can be detected by autocorrelation. This technique allows for a simple instrument for Jovian icy moon sounding, requiring a wire dipole antenna receiver and digitizing electronics with no transmitter. It was found that, under favorable conditions, a dipole antenna receiver using Jovian bursts as a signal of opportunity provided a sensitivity to subsurface oceans comparable to an active radar sounder. 

\begin{figure}[h!]
\includegraphics[width=1.0\linewidth]{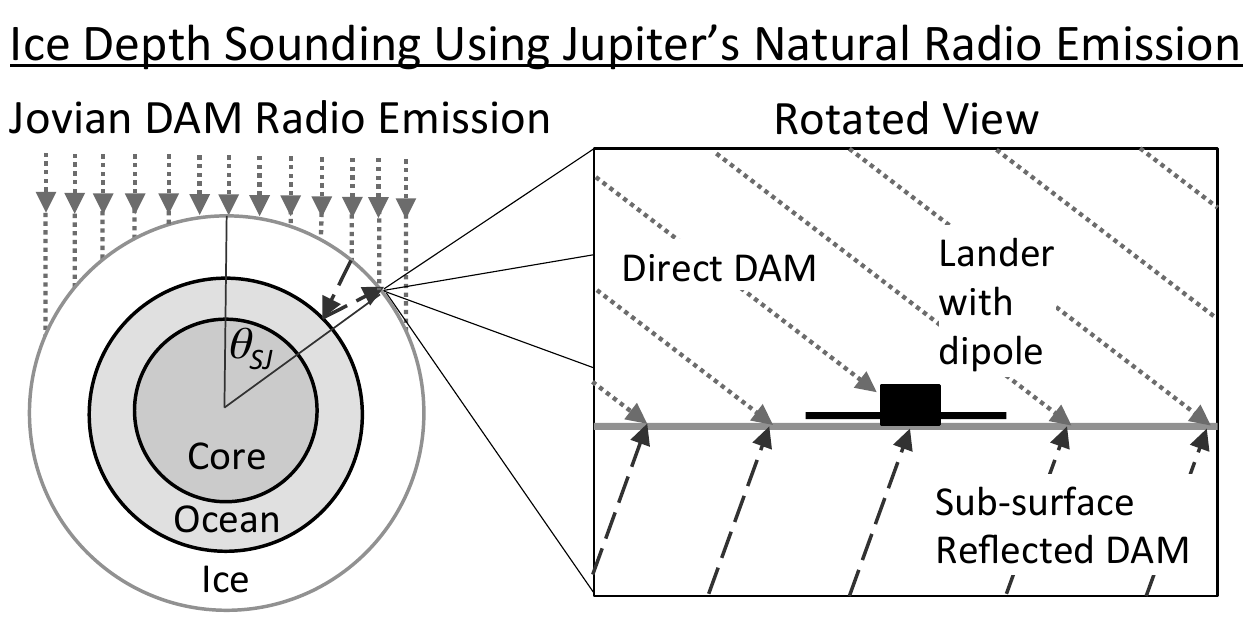}
\caption{Passive detection of subsurface oceans in icy moons using Jupiter's radio emission and its echoes with a lander. The Jovian decametric (DAM) emission is incident on the subjovian side of Europa approximately as a plane wave (gray arrows) and illuminates the ice. The dipole antenna on a lander at an angle $\theta_{SJ}$ from the subjovian point samples the direct radio emission from Jupiter (dotted gray line). The Jovian DAM emission refracts into the ice shell, reflects off the subsurface ocean (arrows with dashed-dotted lines), and propagates to the lander. Correlation of the direct and reflected emission results in a delayed peak traceable to the thickness of the ice shell.} 
\label{fig:LANDER_concept} 
\vskip -0.25cm
\end{figure}

In this paper, we estimate the sensitivity of a dipole antenna receiver and correlator on-board a lander on the subjovian side of Europa to reflections from a subsurface ocean at Europa (see Figure~\ref{fig:LANDER_concept}). In Romero-Wolf et al. 2015 it was found that the main factors limiting the sensitivity were the surface roughness of the ice-atmosphere boundary and the $\lesssim$1-second limit on integration times due to the motion of the orbiter. For a lander these limitations are significantly reduced. The antenna would lie directly on the surface, which eliminates scattering and transmission losses due to surface roughness at the ice-atmosphere interface, a major source of uncertainty and clutter. The receiver is stationary, significantly increasing the possible integration time. 

Recently, Grima et al. 2015 pointed out that Faraday rotation in the ionosphere of Europa would limit the minimum usable frequency of high altitude radar to 4-8~MHz. For a lander, not only does the radio signal traverse the ionosphere in only one direction, reducing dispersion effects, but also the direct and reflected signal both have nearly equal ionospheric dispersion, effectively canceling the effect after correlation. The dominant ionospheric effect will thus be the blockage of radio signals below the plasma frequency, which is $\sim$1~MHz.

Jovian decametric emission is variable but predictable. The strongest bursts come from the Io-driven sources, which depend on the longitude of Jupiter and orbital phase of Io (Bigg 1964; Payan 2014). In Romero-Wolf et al. 2015, it was shown that the Io-driven source is compact enough (Dulk 1970; Carr et al. 1970; Lynch et al. 1976) to be treated effectively as a point source for the purposes of sounding. The main limitation in integration time arises from the limited time of duration of the Io-driven sources on Europa and their motion on the sky. However, since Europa is tidally locked, the location of the Jovian DAM will shift by $\lesssim$ $1^{\circ}$ on the sky. At Europa, the Io-driven bursts are expected to last $\sim$90~minutes on average and for as long as a couple of hours. These bursts occur at predictable times once every two to three days.


We estimate the signal strength of subjovian decametric emission for a lander located at $\theta_{SJ}$ defined as the angle between the lines connecting the center of Europa to the sub-Jovian point and the center of Europa to the position of the lander. The fraction of the Jovian DAM power transmitted into the ice layer is given by $(1-\rho_{ice-atm})$, where $\rho_{ice-atm}$ is the reflection coefficient of the ice-atmosphere interface and depends on the angle of incidence. The path length from the surface entry point to the subsurface ocean at depth $d$ is given by $D=d/\cos\theta_r$, where $\theta_r$ is the refracted angle given by Snell's law. The geometric relations between the angle of incidence, the refracted angle $\theta_r$, and the location of the observer $\theta_{SJ}$ are provided in Romero-Wolf et al. 2015. The signal losses due to propagation and absorption in ice is given by $10^{-\alpha D / 10}$, where $\alpha$ is the two-way attenuation rate in ice given in dB/km. The reflection off the ice-ocean interface contributes a factor of $\rho_{ice-ocn}$, which is expected to be near unity. Given the radio flux density of Jupiter $S_{J}$ incident on the surface of Europa, the flux density $S_{R}$ of the signal transmitted through the ice-atmosphere layer, propagated down to the ice-ocean layer, reflected off this layer, and propagated back through the ice-atmosphere layer is
\begin{equation}
S_{R}=10^{-\alpha D / 10}(1-\rho_{ice-atm})^2\rho_{ice-ocn}S_{J}.
\label{eqn:reflected_flux}
\end{equation}

As discussed in Romero-Wolf et al. 2015, the dominant noise source is, by far, the Jovian decametric emission itself. For a correlation averaged over time $\Delta T$ and bandwidth $\Delta f$, the amplitude signal-to-noise ratio given by $SNR = \sqrt{\Delta T \Delta f S_R/S_J}$, which, combined with Equation~\ref{eqn:reflected_flux}, results in
\begin{equation}
SNR=10^{-\alpha D / 20}(1-\rho_{ice-atm})\sqrt{\rho_{ice-ocn}}\sqrt{\Delta T \Delta f}.
\end{equation}
For a detection threshold signal-to-noise ratio of $SNR_{thr}$, the maximum detectable depth is given by
\begin{equation}
d_{max}=\frac{20}{\alpha}\cos\theta_r\log_{10}\left[(1-\rho_{ice-atm})\sqrt{\rho_{ice-ocn}}\frac{\sqrt{\Delta T \Delta f}}{SNR_{thr}}\right].
\end{equation}
The first factor in the logarithm, due to transmission through the ice and reflection of the subsurface ocean, is near unity. The ice shell two-way attenuation rate $\alpha$ was estimated for Europa by Moore 2000 and Blankenship et al. 2009 to be 1.6 dB/km for pure ice and 16 dB/km for high loss ice. 

We estimate values of the maximum detectable depths with the following assumptions. We use an integration time of 90 minutes, corresponding to the average duration of a Jovian burst. 
The bandwidth of integration is 20~MHz limited by the ionospheric cutoff of 1~MHz on the low end and the magnetic cutoff at 40~MHz, with allowance for all frequencies not being active over the period of the burst. The requirement for detection is that the signal-to-noise ratio be greater than one. These assumptions, in the case of $\theta_{SJ}=0^{\circ}$, where the performance is best, result in a maximum detectable depth of 6.9~km for high loss ice and 69~km for pure ice. We can compare these sensitivities to the current bounds on the ice shell thickness of Europa. The lowest estimates of the ice shell thickness of Europa are $\gtrsim$3~km from Pappalardo 1998 and Turtle \& Pierazzo 2001. The upper bound on the ice shell thickness of Europa is 30~km (Ojakangas and Stevenson 1989). 

The depth resolution for this technique is, at best, given by $c/(2\Delta f)$, where $c$ is the speed of light (see Romero-Wolf et al., 2015). For 20~MHz bandwdith this gives a lower bound on the depth resolution of 7.5~meters. The resolution could be degraded by the non-ideal autocorrelation characteristics of the Jovian burst. However, the autocorrelation function of the Jovian bursts at and around zero delay provides the impulse response function. This allows for the quantification of the resolution from the data itself. It is expected that the resolution is in the order-of-magnitude scale of 10~meters.

We estimate the maximum detectable depth for an active radar instrument on-board a flyby spacecraft and compare it to the sensitivity of a lander-based passive sounder. The received signal strength of an active radar instrument is given by the radar equation
\begin{equation}
\begin{split}
P_{R} = & P_{T}G \frac{ A_{eff}}{4\pi\left(2\left(h+d\right)\right)^2} 
\\
& \times (1-\rho_{atm-ice})^2\rho_{ice-ocn}L_{rough}10^{-\alpha d / 10}
\\
&\times \tau_p \Delta f \Delta T^2 \ \mbox{PRF}^2 
\end{split}
\end{equation}
We assume a high frequency (HF) radar system operating at 9~MHz with transmitter power $P_T=10$~Watts using a dipole antenna with gain $G=1.5$ corresponding to an effective area $A_{eff}=130$~m$^2$. We consider a radar with altitude $h=100$~km. The received power includes propagation losses due to distance and transmission through the ice and reflection off the ice ocean surface. The surface roughness losses are denoted by $L_{rough}$. The system has a compression gain of $\tau_{p}\Delta f$ given by a radar pulse width $\tau_{p}=100$~$\mu$s and bandwidth $\Delta f=1$~MHz. The radar signals with pulse repetition frequency PRF and integration time $\Delta T$ can be added coherently leading to a received power gain of $\Delta T^2 \mbox{PRF}^2$. For a Europa flyby mission, we assume the integration time $\Delta T$ is determined by the spacecraft velocity of $\sim 4$~km/s relative to ground and Fresnel zone of $\sim$2.6~km to give $\Delta T\sim0.6$~s. We consider a pulse repetition frequency PRF=~250~Hz. For an active HF radar on the antijovian side, which is occulted from Jupiter's DAM emissions, the dominant source of noise is due to the galaxy. The galactic noise at $9$~MHz has a flux $\Phi_{gal}\sim1.3\times 10^{-19}$~W/m$^2$/Hz. The noise power due to the galactic background is given by $P_{noise}=\frac{1}{2} \ \Phi_{Gal} \ A_{eff} \ \Delta f \ \Delta T \ \mbox{PRF}$, which includes the effective area of the antenna, receiver bandwidth, and the incoherent amplitude stacking of the noise.
For the system described above, this results in noise power $P_{noise}\sim7.4\times10^{-10}$~W. With these parameters, the SNR of the coherently summed return radar pulse amplitudes is given by an amplitude $SNR=\sqrt{P_{R}/P_{noise}}\sim2.4\times 10^{3} \sqrt{L_{rough}} 10^{-\alpha d / 20}$. 

With the values given above, we estimate the maximum detectable ocean depth for a flyby HF radar. In the optimistic case where there are negligible losses from surface roughness, we estimate $d_{max}\sim$4.0~km for high loss ice and $d_{max}\sim$40 km for pure ice.

\begin{figure}[t!]
\centering
\vskip -0.3cm
\includegraphics[width=1.0\linewidth]{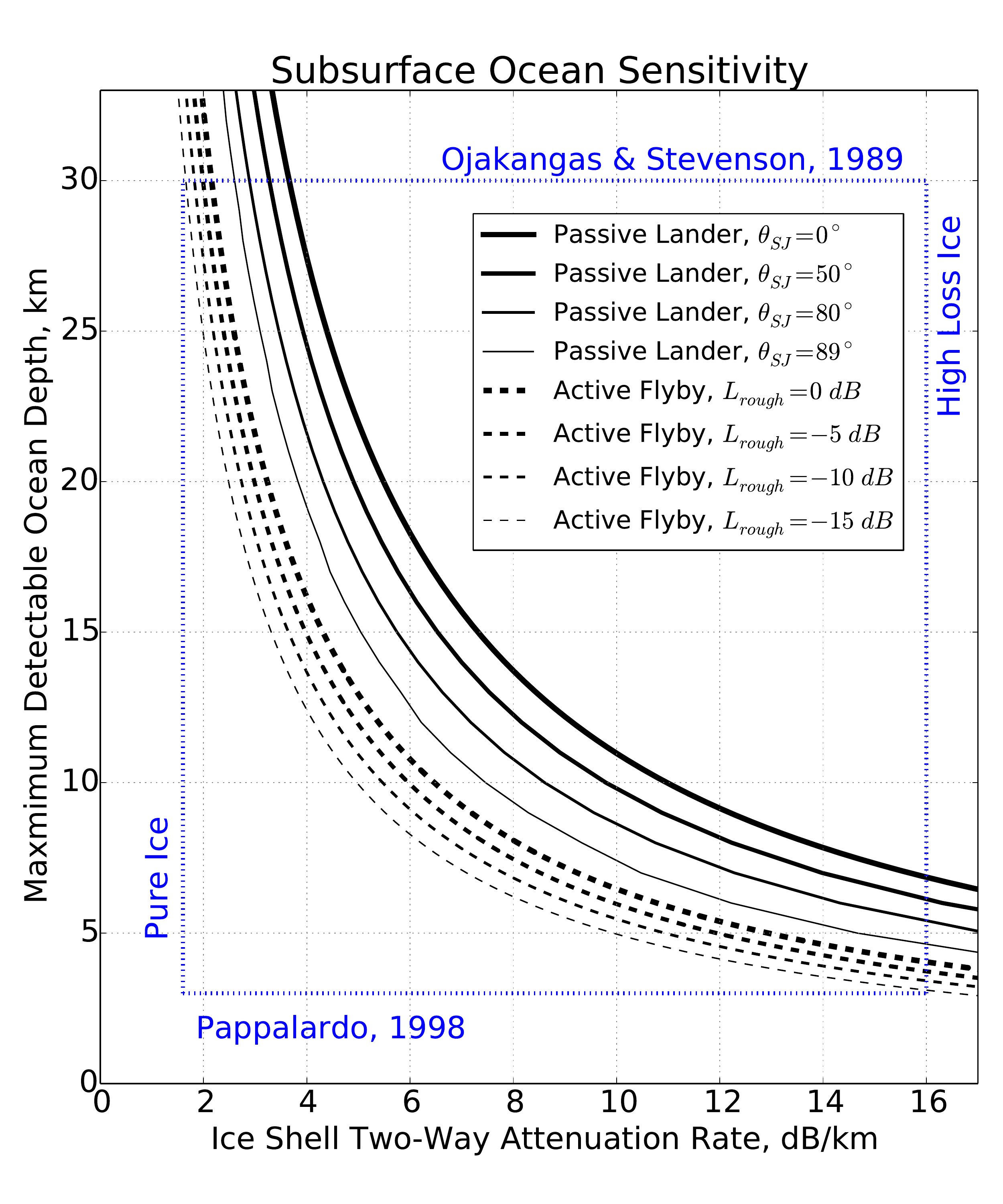}
\vskip -0.5cm
\caption{Sensitivity of a lander-based passive sounder compared to models and predictions of the Europan ice-shell thickness and two-way attenuation rate. The region below each line consists of combinations of two-way attenuation rate and ice shell thickness that would produce a detection of a subsurface ocean. See text for details on the parameters producing this curve. Expected sensitivities are included at the subjovian point $\theta_{SJ}=0^{\circ}$ and for surface distances to the subjovian point of $\theta_{SJ}=50^{\circ}$, $80^{\circ}$, and $89^{\circ}$, where signal losses occur due to decreased transmission coefficients and increased path length through the ice. The lander passive sounder is compared to an active flyby radar (black dashed line) at 100~km altitude (see text for the parameters used in the estimate). The black dashed line assumes a smooth ice-atmosphere surface $L_{rough}=0 \ dB$, which is the most optimistic case. The effect of the ice-atmosphere surface roughness on the sensitivity is illustrated by the dashed lines assuming $L_{rough}$ corresponding to 5,~10,~and~15~dB losses. All sensitivity estimates presented here assume a smooth ice-ocean interface. The lander is sensitive to a larger portion of the model parameter space bounded by the blue lines. The horizontal blue lines mark the range of the subsurface ocean depths from Pappalardo, 1998; Hand \& Chyba 2007; Schenk 2002; and Ojakangas \& Stevenson 1998. The vertical lines mark the range of the attenuation rates encompassing pure ice and high loss ice from Moore 2000 and Blankenship 2009.} 
\label{fig:results} 
\vskip -0.25cm
\end{figure}

Figure~\ref{fig:results} compares the sensitivity of a passive sounder on a lander to the estimates derived for active sounding on a flyby. Additional considerations for signal loss are included in both cases. The passive technique on a lander requires, at the very least, that Jupiter be in view of the instrument. The reflection path is optimal if it is incident normal to the ice surface, which coincides with the subjovian point $\theta_{SJ}=0^{\circ}$, where the path length in the ice would be minimized and the Fresnel transmission coefficient is highest. At locations away from the subjovian point, the transmission coefficient is decreased and the propagation path length in the ice is increased, resulting in more attenuation of the signal. Estimates are provided for surface distances of $\theta_{SJ}=50^{\circ}$, $80^{\circ}$, and $89^{\circ}$ away from the subjovian point. For $\theta_{SJ}>89^{\circ}$, not shown in Figure~\ref{fig:results}, the maximum detectable depth quickly falls below the active flyby radar capabilities. For $\theta_{SJ}>90^{\circ}$, the direct Jovian emission is occulted by the ice shell and this technique cannot be applied.

To illustrate the effects of losses due to surface roughness on the maximum detectable ocean depth, we have also included estimates of the active flyby assuming losses $L_{rough}$ of 5, 10, and 15 dB. While surface roughness can degrade the sensitivity of a flyby radar, the lander system is not sensitive to these effects since the antenna would lie on the atmosphere-ice surface. In addition, the lander-based passive sounder offers significant advantages due to the increased integration time and bandwidth. 

These sensitivities are compared to the current constraints on the ice shell of Europa. The bounds shown in Figure~\ref{fig:results} encompass inferences of the ice shell thickness (Pappalardo 1998; Turtle \& Pierazzo 2001; Hand \& Chyba 2007; Schenk 2002) and predictions of the attenuation rate (Moore 2000 and Blankenship et al. 2009). The improvements offered by a lander instrument have a significant impact on current constraints of the ice shell thickness and attenuation rate. A high sensitivity measurement at a single point in the subjovian side of Europa could be complementary to the larger coverage available to an active flyby radar sounding campaign.


The passive sounder described here could have a low resource implementation on a lander. The Io-driven flux of Jupiter on Europa is in excess of $10^{-15}$~W/m$^2$/Hz. Based on the theory of electrically short antennas, as applied to radio astronomy (see Ellingson 2005), a 1~meter length dipole made of standard measuring tape material would produce a signal that is well above the thermal noise of a receiver. The signal could be sampled 80 MHz with a low-resolution digitizer and fed to an on-board correlator to produce the time-averaged waveforms. The mass and power requirements of such a system could be well within the constraints of a lander on Europa.

The required data rate can be estimated based on the bandwidth and observation time available. For a subsurface reflection below $\sim$ 60~km of ice, allowing for ample margin over the 30~km upper bound, and an index of refraction for ice of 1.7, the round-trip-time for a radio signal is $\sim$~0.7~ms. If digitizing at 80~MHz, this results in 56,000 samples per correlation. Assuming 12-bit samples for an averaged correlation waveform\footnote{Note that this is the resolution of the averaged correlated waveform. The digitizer sampling the raw data could operate with as low as 1-bit resolution.}, this results in $\sim$ 672 kilobytes. Although the correlation can be averaged over 90 minutes, in practice, it would be recommended that integrations be averaged over smaller intervals and then averaged again in post-downlink analysis. If 10 minute intervals are taken, then the data size for the full duration of the Jovian burst increases to 6 megabytes.

Another important concern for a Europa lander is the radiation environment. Energetic charged particles preferentially strike in the trailing hemisphere with the highest energies ($\sim$25~MeV) near the equatorial region (Patterson, Paranicas, \& Procter 2005). Since Europa is tidally locked, the great circle dividing the subjovian and antijovian hemispheres is orthogonal to the great circle dividing the leading and trailing hemispheres with intersecting points at the poles. A passive sounder on a lander could therefore operate in the leading-subjovian quadrant of Europa.

A single lander could provide observations to improve the constraints in the ice shell thickness of Europa. A positive detection could disambiguate whether the differentiation in the ice layer of Europa is due to a liquid subsurface layer or warm convective ice. A single point measurement in the subjovian side of Europa with this technique could be a powerful complement to an active flyby radar sounding campaign. It is conceivable that a number of small probes equipped with deployable tape-measure dipoles and low-power electronics could be placed on various points along the surfaces of Jovian icy moons to map out the depths of their subsurface oceans.


{\it Acknowledgements:} This research was carried out at the Jet Propulsion Laboratory, California Institute of Technology, under a contract with the National Aeronautics and Space Administration. Copyright 2016 California Institute of Technology. Government sponsorship acknowledged.

%
%
%



\begin{thebibliography}{9}

%

\bibitem{key} Bigg, E.K. (1964), ``Influence of satellite Io on Jupiter’s decametric emission", Nature, 203, 1008-1009.

\bibitem{key} Blankenship, D.~D., Young, D.~A., Moore, W.~B., and Moore, J.~C. (2009), ``Radar Sounding of Europa's Subsurface Properties and Processes: The View from Earth.", article in ``Europa", edited by Pappalardo,~R.~T.., McKinnon~W.~B, Khurana~K.~K., University of Arizona Press, 631-653

\bibitem{key} Carr, T.D, et al. (1970), ``Very long baseline interferometry of Jupiter at 18 MHz", Radio Science, 5, 1223-1226

\bibitem{key} Dulk, G.A. (1970), ``Characteristics of Jupiter's Decametric Radio Source Measured with Arc-Second Resolution", Astrophysical Journal, 159, 671

\bibitem{key} Ellingson, S.~W., (2005), ``Antennas for the Next Generation of Low-Frequency Radio Telescopes", IEEE Transaction on Antennas and Propagation, 53, 2480-2489.

\bibitem{key} Grima, C., Blankenship, D.~D., Schroeder, D.~M., (2015), ``Radar signal propagation through the ionosphere of Europa", Planetary and Space Science, 117, 421-428

\bibitem{key} Hand, K. P., and C. F. Chyba (2007), ``Empirical constraint on the salinity of the Europan ocean and implications for a thin ice shell", Icarus, 189, 424-438

\bibitem{key} Lynch, M.A., Carr, T.D., May, J. (1976), ``VLBI Measurements of Jovian S. Bursts", Astrophysical Journal, 207, 325-328

\bibitem{key} Moore, J.~C., (2000), ``Models of Radar Absorption in Europan Ice", Icarus, 147, 292-300

\bibitem{key} Ojakangas, G.W., Stevenson, D.J. (1989), ``Thermal state of an ice shell on Europa", Icarus, 81, 220-241

\bibitem{key} Pappalardo, R. T., et al. (1998), ``Geological evidence for solid-state convection in Europa's ice shell", Nature, 391, 365-368

\bibitem{key} Patterson, G.~W., Paranicas, C., and Procter, L.~M. (2005) ``Characterizing electron bombardment of Europa’s surface by location and depth", Icarus, 220, 286-290.

\bibitem{key} Payan, A.P et al. (2014) ``Effect of plasma torus density variations on the morphology and brightness of the Io footprint", J. Geophys. Res., 119, 3641-3649.

\bibitem{key} Romero-Wolf, A., Vance, S., Maiwald, F., Heggy, E., Ries, P., Liewer, K., (2015), ``A passive probe for subsurface oceans and liquid water in Jupiter’s icy moons", Icarus, 248, 463-477

\bibitem{key} Schenk, P.M. (2002), ``Thickness constraints on the icy shells of the galilean satellites from a comparison of crater shapes", Nature, 417, 419-421 

\bibitem{key} Turtle, E.P. and Pierazzo, E. (2001), ``Thickness of a Europan Ice Shell from Impact Crater Simulations", Science, 294, 1326-1328 




%
%
%
%
%
%
%
%
%
%
%
%
%
%
%
%
%
%
%
%
%
%
%
%
%
%
%
%
%
%
%
%
%
%
%
%
%
%
%
%
%
%
%
%
%
%
%
%
%
%
%
\end{thebibliography}
\end{document}